\documentclass[conference]{IEEEtran}

\usepackage[utf8]{inputenc}
\usepackage{cite}
\usepackage{amsmath,amssymb,amsfonts}
\usepackage{graphicx}
\usepackage{textcomp}
\usepackage{xcolor}
\usepackage{url}
\usepackage{hyperref}
\usepackage{booktabs}
\usepackage[caption=false]{subfig}
\usepackage{placeins}
\newcommand{\ket}[1]{|#1\rangle}

\begin{document}

\title{QuBridge: Layer-wise Fidelity Decomposition\\in Quantum Computation Pipeline}

\author{
\IEEEauthorblockN{Kisho Sotokawa\textsuperscript{1}, Hideaki Kawaguchi\textsuperscript{1}, Shin Nishio\textsuperscript{1,2}, and Takahiko Satoh\textsuperscript{3}}
\IEEEauthorblockA{
\textsuperscript{1}\textit{Graduate School of Science and Technology, Keio University, Yokohama, Kanagawa 223-8522, Japan}\\
\textsuperscript{2}\textit{Department of Physics \& Astronomy, University College London, London, WC1E 6BT, United Kingdom}\\
\textsuperscript{3}\textit{Faculty of Science and Technology, Keio University, Yokohama, Kanagawa 223-8522, Japan}\\
Corresponding author: Kisho Sotokawa (e-mail: kishousotokawa@keio.jp)
}
}

\maketitle

\begin{abstract}
Running a quantum circuit on current hardware involves a sequence of engineering decisions, each with tunable parameters and distinct error characteristics.
Existing tools optimize each decision in isolation, leaving practitioners unable to determine how much each decision contributes to final output quality.
We present QuBridge, a pipeline analysis tool that decomposes quantum computation into three decision layers and measures each layer's fidelity contribution through progressive ablation and isolation experiments.
Applied to quantum teleportation under IBM-calibrated noise models, the framework surfaces three phenomena that end-to-end measurement obscures.
Qubit selection narrows the worst-case fidelity band from $11.8\%$ to under $2\,\%$ with downstream layers held fixed, without changing the peak.
Per-gate pulse-shape assignment adds a $+0.9\%$ residual gain whose attributed magnitude depends on upstream layout.
Error-detection encoding is not uniformly advantageous, and its conditional benefit emerges for input states whose dominant error channel is detectable by the chosen code.
QuBridge operates on cached calibration data without requiring live hardware access.
\end{abstract}

\begin{figure*}[!tb]
\centering
\includegraphics[width=\textwidth]{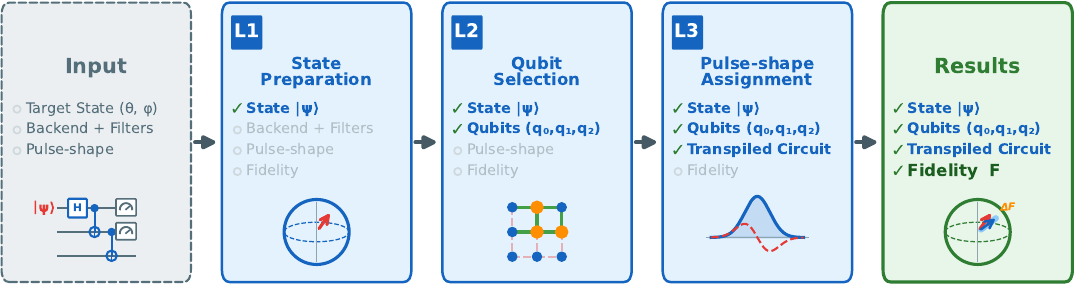}
\caption{QuBridge pipeline architecture. Three decision layers plus Results decompose quantum computation into separately tunable layers. Each layer progressively accumulates pipeline outputs (checked items), culminating in the final fidelity $F$ at Results. Both physical (3-qubit) and error-detected (6-qubit) modes share this structure.}
\label{fig:pipeline}
\end{figure*}


\section{Introduction}
\label{sec:introduction}

Executing a quantum circuit on current hardware requires a chain
of engineering decisions. The practitioner chooses which
backend to target, which physical qubits to allocate, how
transpilation maps logical operations to the device's native
gate set, and how to shape the control pulses that
drive each gate.
On noisy intermediate-scale quantum (NISQ)
devices~\cite{preskill2018quantum}, each decision constrains downstream
options and introduces its own error channel.
However, existing work evaluates these decisions separately.
Noise-aware mapping strategies~\cite{murali2019noiseadaptive,tannu2019noiseadaptive} assess
layout quality without accounting for subsequent transpilation. Pulse
optimization is studied independently of qubit
placement~\cite{gambetta2011leakage}.
When output fidelity changes, practitioners cannot determine which layer
caused the gain or regression, nor whether an upstream improvement
survives downstream transformations at all.

This pipeline structure marks the boundaries where theoretical
equivalence breaks under physical constraints. Mathematically
equivalent input states, topologically equivalent qubit
placements, and unitary-equivalent gate decompositions yield
different hardware outcomes through state-dependent
dephasing, spatial noise heterogeneity, and spectral leakage
respectively.

Current tools do not expose these boundaries.
Circuit simulators such as Quirk~\cite{gidney2017quirk} enable gate-level
exploration without hardware context.
Quantum compilers including Qiskit~\cite{javadi2024quantum},
\texttt{tket}~\cite{sivarajah2021tket}, and
BQSKit~\cite{younis2021bqskit} optimize individual compilation passes but
return only a final compiled circuit. Per-pass fidelity contributions are
not reported.
Visualization platforms such as IBM Quantum
Composer~\cite{ibm2016composer} and Quantum
Flytrap~\cite{jankiewicz2022quantum} address single stages in isolation.
The quantum software engineering community~\cite{zhao2020quantum,
piattini2020quantum} has begun formalizing testing and quality-assurance
methodologies~\cite{wang2021qdiff}, but no existing tool measures whether
an improvement at one decision, such as selecting lower-error qubits,
actually survives subsequent transpilation and noise, or is silently
absorbed by later decisions.

We address this attribution problem through layer-wise fidelity
decomposition, the attribution of end-to-end fidelity change to individual
pipeline layers, each traceable to a dominant hardware-level error mechanism.
We operationalize this through two complementary protocols.
Progressive ablation optimizes one layer while randomizing the rest,
measuring that layer's contribution as a reduction in fidelity band width.
Isolation experiments fix all other layers and vary a single parameter set
to expose one layer's sensitivity profile.
Quantum teleportation~\cite{bennett1993teleporting} serves as the primary
benchmark because it exposes layer-wise structure clearly. The same three
theoretical breakpoints described above appear at distinct, identifiable
points in the protocol, and its compact circuit size makes noise
contributions attributable rather than conflated.

We present QuBridge, a pipeline analysis tool that decomposes quantum
computation into three decision layers (state preparation, qubit
selection, and pulse-shape assignment) plus a results layer, and measures each
layer's fidelity contribution.
We validate on physical (three-qubit) and error-detected (six-qubit)
quantum teleportation using noise-model simulation calibrated to IBM
Torino (133-qubit Heron architecture).

Rather than automating pipeline optimization, QuBridge is
designed as an exploratory environment where practitioners
can observe how decisions at each layer, including
error-detection encoding, shape end-to-end fidelity. The
tool's value lies in making improvement mechanisms visible
rather than in producing optimal circuits automatically.
Our contributions are threefold.

\begin{enumerate}
\item We present QuBridge, a pipeline analysis tool that 
decomposes quantum execution into three decision layers and 
reports per-layer fidelity contributions through progressive 
ablation and isolation experiments, operating on cached 
calibration data without live hardware access.

\item We ground layer-wise fidelity decomposition in the 
theory-hardware boundary, where each layer is motivated by a  
point where theoretically equivalent decisions yield physically 
distinct outcomes, making per-layer attribution operationally meaningful.

\item Applied to physical and error-detected teleportation,
the framework surfaces three phenomena that end-to-end
measurement obscures. Qubit selection acts on the worst-case
rather than the peak fidelity. Per-layer contributions are
context-dependent with respect to upstream state.
Error-detection encoding is not uniformly advantageous, and
its conditional benefit emerges for input states whose
dominant error channel is detectable by the chosen code.
\end{enumerate}

\section{Related Work}
\label{sec:related}

\textbf{Tools for quantum execution pipelines.} 
Quantum compilers including Qiskit~\cite{javadi2024quantum}, 
tket~\cite{sivarajah2021tket}, and 
BQSKit~\cite{younis2021bqskit} optimize individual 
compilation passes and return a final compiled circuit. 
Qiskit's PassManager exposes per-pass callbacks for 
intermediate inspection, but at the granularity of 
compilation passes rather than decision layers that map to 
distinct physical error mechanisms. Noise-aware mapping 
strategies~\cite{murali2019noiseadaptive, 
tannu2019noiseadaptive} and VF2 subgraph 
matching~\cite{cordella2004vf2} improve qubit layout 
selection but evaluate this stage in isolation. 
Visualization platforms such as Quirk~\cite{gidney2017quirk}, 
IBM Quantum Composer~\cite{ibm2016composer}, Quantum 
Flytrap~\cite{jankiewicz2022quantum}, and QuTiP Virtual 
Lab~\cite{qutipvlab} address single stages without exposing 
per-layer fidelity attribution. Recent quantum software 
debugging tools~\cite{wang2021quito} target test generation 
and assertion checking, and QuBridge addresses the 
orthogonal problem of attributing fidelity changes to 
pipeline decisions.

\textbf{Attribution and decomposition.} Attributing observed
behavior to individual components is a recurring pattern in
software engineering.
Delta debugging~\cite{zeller2002simplifying} progressively isolates
the change responsible for a regression by ablating one
factor at a time, and spectrum-based fault
localization~\cite{jones2005empirical} attributes test
failures to specific code locations through similar
ablation-style reasoning. 
While testing and quality assurance for quantum
software is being formalized~\cite{zhao2020quantum,
piattini2020quantum, wang2021qdiff}, per-layer fidelity
attribution remains unaddressed for the quantum execution
pipeline.
QuBridge applies the progressive-ablation pattern to the
quantum execution pipeline, where each layer corresponds to
a dominant hardware-level error mechanism rather than an arbitrary
software boundary.
\section{System Design}
\label{sec:system}

\subsection{Pipeline Architecture}

\begin{figure*}[!tb]
\centering
\subfloat[Waterfall view.]
{\includegraphics[height=5.7cm]{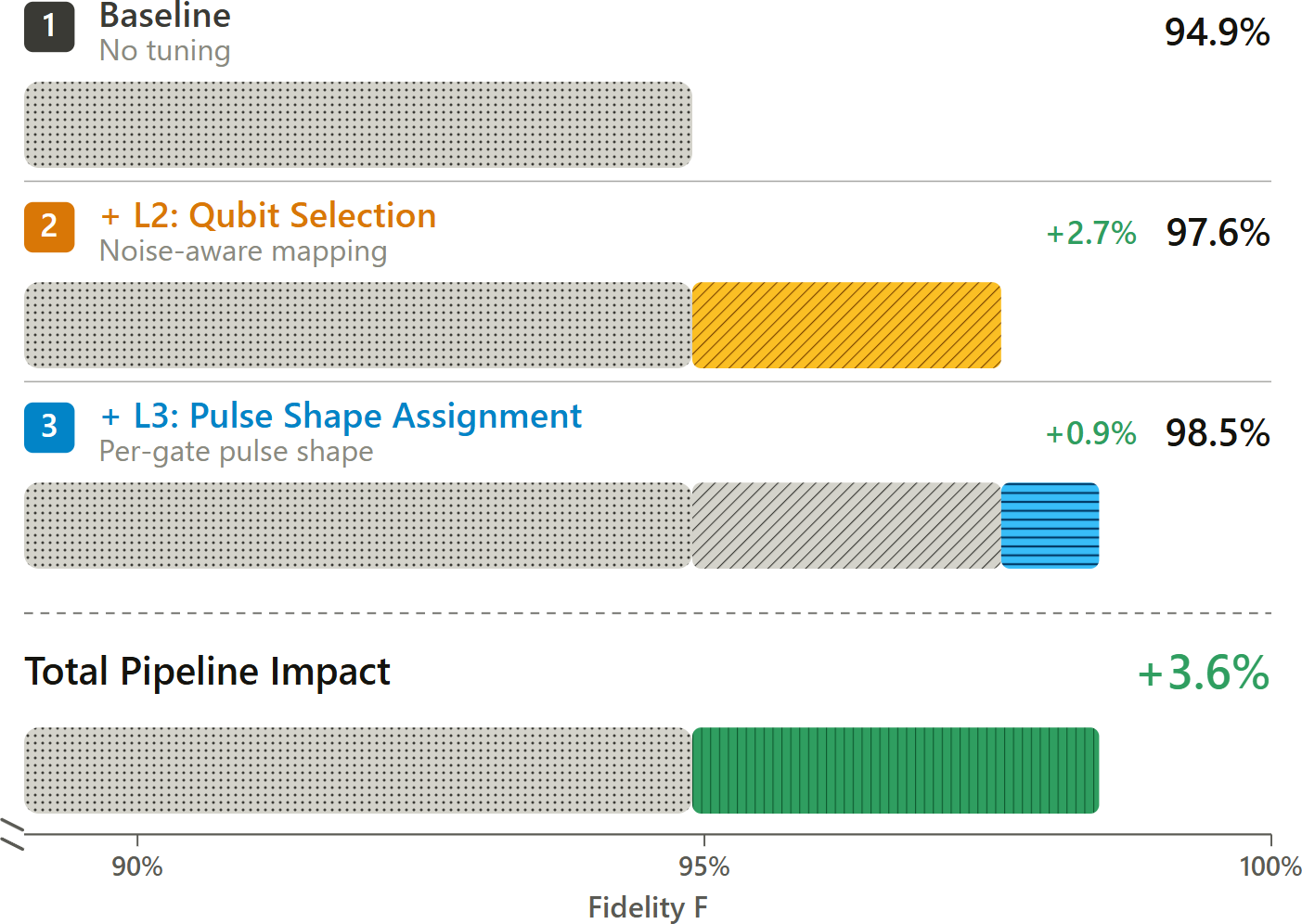}\label{fig:results_ui}}\hspace{2mm}
\subfloat[Progressive ablation result.]{\includegraphics[height=5.7cm]{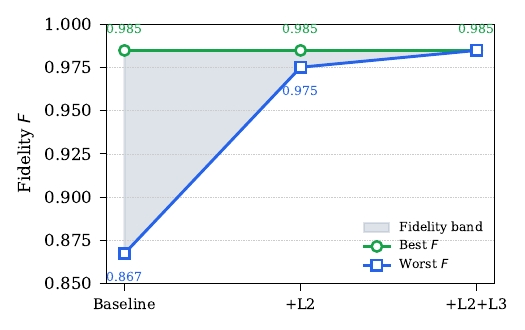}\label{fig:results_band}}
\caption{Results and Pipeline Decomposition. (a)~Illustrative rendering of the Waterfall view, which decomposes the $+3.6\%$ pipeline gain into Baseline, +L2 (Qubit Selection), and +L3 (Pulse-shape Assignment). Numbered badges (1, 2, 3) act as live Fidelity Actions buttons that add each data point. (b)~Band width narrows as L2 and L3 are progressively optimized, confirming L2 provides the largest band narrowing.}
\label{fig:results}
\end{figure*}

QuBridge decomposes quantum computation into three decision layers (L1--L3) plus Results (Fig.~\ref{fig:pipeline}). Each decision layer exposes tunable parameters whose downstream fidelity effects are measured in Section~\ref{sec:evaluation}. Results is a measurement stage that reports $F$ and per-layer contributions, not a decision layer.

The system supports two teleportation modes sharing the same pipeline,  \textbf{physical} (three-qubit) and \textbf{error-detected} (six-qubit with $[[2,1,1]]$ repetition code encoding).

\textbf{State preparation (L1).} The input state $\ket{\psi} = \cos(\theta/2)\ket{0} + e^{i\phi}\sin(\theta/2)\ket{1}$ is prepared using $R_y(\theta)$ and $R_z(\phi)$ gates, parameterized via interactive Bloch sphere sliders.

\textbf{Qubit selection (L2).} L2 selects the physical qubits and coupling-map edges on which the circuit executes. Mode-specific mechanisms (threshold filtering, VF2 subgraph matching) are detailed in \S\ref{sec:l2} and \S\ref{sec:logical}.

\textbf{Pulse-shape assignment (L3).} L3 assigns a pulse envelope (Square, Gaussian Square, or derivative 
removal by adiabatic gate (DRAG)) to each native gate type in the transpiled circuit. Duration and amplitude are fixed by calibration, so L3 controls only the envelope shape.

\textbf{Results.} In the current implementation, we visualize the preparation and output of single-qubit states via Bloch sphere rendering, suited to the teleportation workload demonstrated in this paper. The Results layer is modular, and alternative renderers such as expectation-value plots or measurement histograms can be substituted for circuits with different output structures.

To make per-layer contributions observable during interactive use, Results presents a \textbf{Layer Fidelity Waterfall} (Fig.~\ref{fig:results_ui}) that decomposes the total fidelity improvement into per-layer bars,  a baseline (default qubits, no pulse-shape assignment) plus the incremental gain from each pipeline decision (L2 noise-aware qubit selection, L3 per-gate pulse-shape assignment).

This paper empirically focuses on L2, L3, and one
encoding-mediated interaction case. L1 is exposed as an
interactive control surface rather than systematically swept,
and its quantitative contribution is not reported here.

\subsection{Error-Detection Encoding}
\textbf{Encoding as a meta-decision.} Encoding is a 
meta-decision that reshapes the L2 layout landscape rather 
than acting as a separate decision layer.

The error-detected mode uses a $[[2,1,1]]$ repetition code, with encoded states $\ket{0_L}{=}\ket{00}$ and $\ket{1_L}{=}\ket{11}$, and logical operators $X_L{=}X{\otimes}X$ and $Z_L{=}Z{\otimes}I$. This yields a six-qubit circuit (Alice qubits~0--1, Mediator 2--3, Bob 4--5) with 8~two-qubit gates, amounting to $4.0\times$ the physical circuit's 2~two-qubit gates. The code detects single bit-flip errors by flagging measurement outcomes outside $\{00,11\}$, enabling syndrome-based post-selection without ancilla overhead. This code is chosen because its compact circuit graph maps to many adjacent configurations on current hardware, enabling layouts that minimize routing overhead and preserve the detection margin.

\section{Evaluation}
\label{sec:evaluation}

\begin{figure*}[!tb]
\centering
\subfloat[Filter schematic.]{\includegraphics[height=6.1cm]{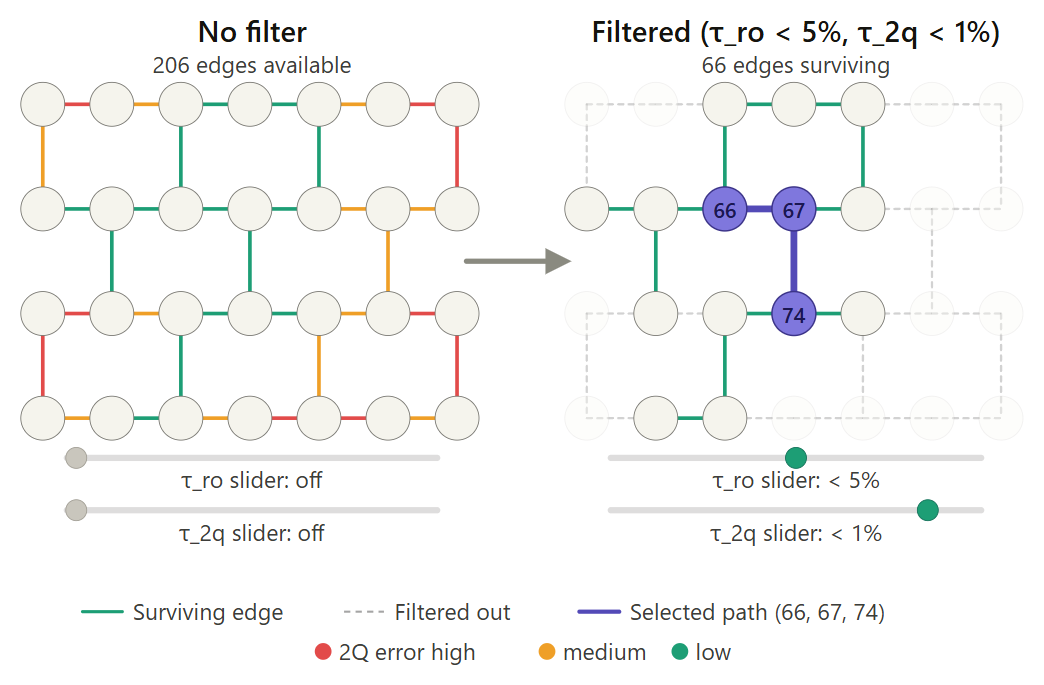}\label{fig:l2_ui}}\hspace{2mm}
\subfloat[Filter cascade result.]{\includegraphics[height=6.1cm]{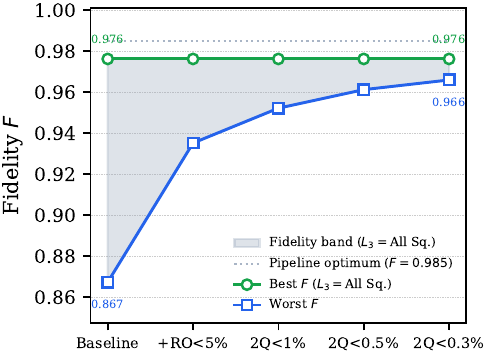}\label{fig:l2_result}}
\caption{L2 Qubit Selection. (a)~Illustrative rendering of the filter panel. Edges are colour-coded by 2Q gate error (red high, green low). Sliders $\tau_\text{ro}$ and $\tau_\text{2q}$ progressively prune high-error edges, narrowing the edge pool from 206 to 66. The surviving subgraph (highlighted triplet $(66,67,74)$) is what the user selects from. Edge count and the corresponding number of three-qubit candidate paths through the filtered subgraph are reported separately in Table~\ref{tab:e1g}. (b)~The fidelity band narrows ($11.8\%{\to}1.9\%$) across the cumulative filter cascade. Best $F{=}0.985$ is the pipeline optimum, reached in Fig.~\ref{fig:l3_result} via L3 tuning.}
\label{fig:l2}
\end{figure*}

We evaluate layer-wise fidelity decomposition through four experiments. Progressive ablation tests whether per-layer contributions can be isolated (\S\ref{sec:trace}). An L2 isolation sweep identifies why qubit selection produces the largest band narrowing (\S\ref{sec:l2}). An L3 isolation experiment quantifies the effect of per-gate pulse-shape assignment (\S\ref{sec:l3}). Extension to error-detection encoding tests whether the framework generalizes (\S\ref{sec:logical}). The primary metric is state fidelity $F = \langle\psi|\rho|\psi\rangle$, which directly measures teleportation quality. Across the progressive ablation experiments (\S\ref{sec:trace}, \S\ref{sec:logical}), we quantify a layer's \emph{fidelity contribution} as the band narrowing it induces. We define $C_\ell = \mathrm{Band}_{\text{before fixing }\ell} - \mathrm{Band}_{\text{after}}$, where $\mathrm{Band} = F_{\text{best}} - F_{\text{worst}}$ is the width of the fidelity distribution over pipeline configurations. Throughout this section, Band is operationalized as the gap from the worst surviving configuration in a given setting to the best observed pipeline configuration $F = 0.985$, which serves as a practitioner-facing fixed reference.
We use band narrowing as the contribution metric because the
practitioner-facing question is not only how high fidelity
can be made, but how safely poor configurations can be
excluded.

All fidelity values are computed via density-matrix simulation under a noise model built from cached calibration data of IBM Torino (133 qubits, Heron architecture, calibration snapshot 2026-01-16). Each gate's error channel is constructed as a composition of thermal relaxation ($T_1$, $T_2$) and a sparse Pauli-Lindblad residual term~\cite{vandenBerg2023,malekakhlagh2025}, with Pauli generators weighted by decoherence fractions anchored on randomized-benchmarking rates. Readout is modeled as a calibrated asymmetric confusion matrix. Circuits are transpiled with Qiskit~1.3.1 at optimization level~3, and simulation uses Qiskit Aer~0.16.0~\cite{qiskitaer}. Density-matrix simulation is deterministic, so per-configuration repeats in the tables ensure logging-schema consistency across experiments rather than providing statistical noise estimates. Unless stated otherwise, experiments use the stress-test state $\ket{+}$, which is maximally sensitive to the dephasing channel that dominates at our circuit depth. \S\ref{sec:logical} extends to $\ket{1}$ to expose state-dependent behavior under $T_1$ decay.

For interpretability, the progressive ablation baseline fixes
L3 to the uniform All-Square setting while varying L2 across
random configurations. We then expose the residual L3
sensitivity after fixing the best L2 path. This layered
construction, rather than strict simultaneous randomization,
is what Table~\ref{tab:e1h} reports.

\subsection{End-to-End Pipeline Trace}
\label{sec:trace}


We first test whether each layer's fidelity contribution can be 
isolated when the others are randomized, then trace how QuBridge 
surfaces this decomposition to a practitioner.

\textit{Diagnostic vignette.} A practitioner prepares $\ket{+}$
for teleportation on IBM Torino.
Starting from a topology-valid but not noise-aware qubit path
and a uniform Square pulse-shape, she observes $F = 0.949$.
Activating L2 with noise-aware mapping selects the surviving
path $(66, 67, 74)$ and raises fidelity to $F = 0.976$, and
activating L3 with per-gate pulse-shape assignment further
raises it to $F = 0.985$. The Waterfall (Fig.~\ref{fig:results_ui}) decomposes the
cumulative $+3.6\%$ improvement into per-layer bars live.

Without QuBridge, Qiskit's \texttt{transpile()} would have
returned only the final compiled circuit; the practitioner
would observe $F = 0.985$ from end-to-end measurement, with
no visibility into how L2 and L3 each contributed to the
$+3.6\%$ gain.

\textit{Systematic validation.} To confirm that the vignette 
reflects a reproducible decomposition rather than a favorable path, 
we sweep random configurations with one layer progressively fixed 
(Fig.~\ref{fig:results_band}, Table~\ref{tab:e1h}). The fidelity
band narrows from $11.8\%$ (L3 fixed at All Square,
L2 randomized) to $1.0\%$ once L2 is fixed to the noise-aware path. The
residual gap is then closed by L3 per-gate assignment,
which deterministically selects the per-gate optimum
($F=0.985$). Best $F$ pins at $0.985$
once L2 is fixed, confirming that L2 determines the reachable
ceiling while L3 closes the residual worst-case gap.
This collapse from the $11.8\%$ baseline band is the
framework's quantitative signature. Per-layer contributions are 
separable, and the tool renders that separability visible at 
interaction time.





\begin{table}[t]
\centering
\caption{Pipeline Band Narrowing as L2 and L3 are Progressively Optimized}
\label{tab:e1h}
\begin{tabular}{lrcr}
\toprule
Layer & $n$ & Worst $F$ & Band \\
\midrule
Baseline (all L2 paths, $L_3{=}$All Sq.) & 184 & 0.867 & 11.8\% \\
+L2 (fix qubits, $L_3$ varies) & 4 & 0.975 & 1.0\% \\
+L2+L3 (fix qubits, $L_3{=}$per-gate opt.) & 1 & 0.985 & 0.0\% \\
\bottomrule
\end{tabular}
\\[2pt]
{\footnotesize\raggedright Each row sweeps the layers not yet fixed (L2 paths in row 1; L3 shapes in row 2; row 3 fixes both), where $n$ counts the distinct configurations and each is averaged over three repeats. The fixed L2 path is $(66, 67, 74)$. The collapse of the band to zero once both layers are fixed is the framework's quantitative signature, with the reachable ceiling pinned at $F = 0.985$.\par}
\end{table}

\begin{figure*}[!t]
\centering
\subfloat[Pulse-shape comparison.]{\includegraphics[height=6.4cm]{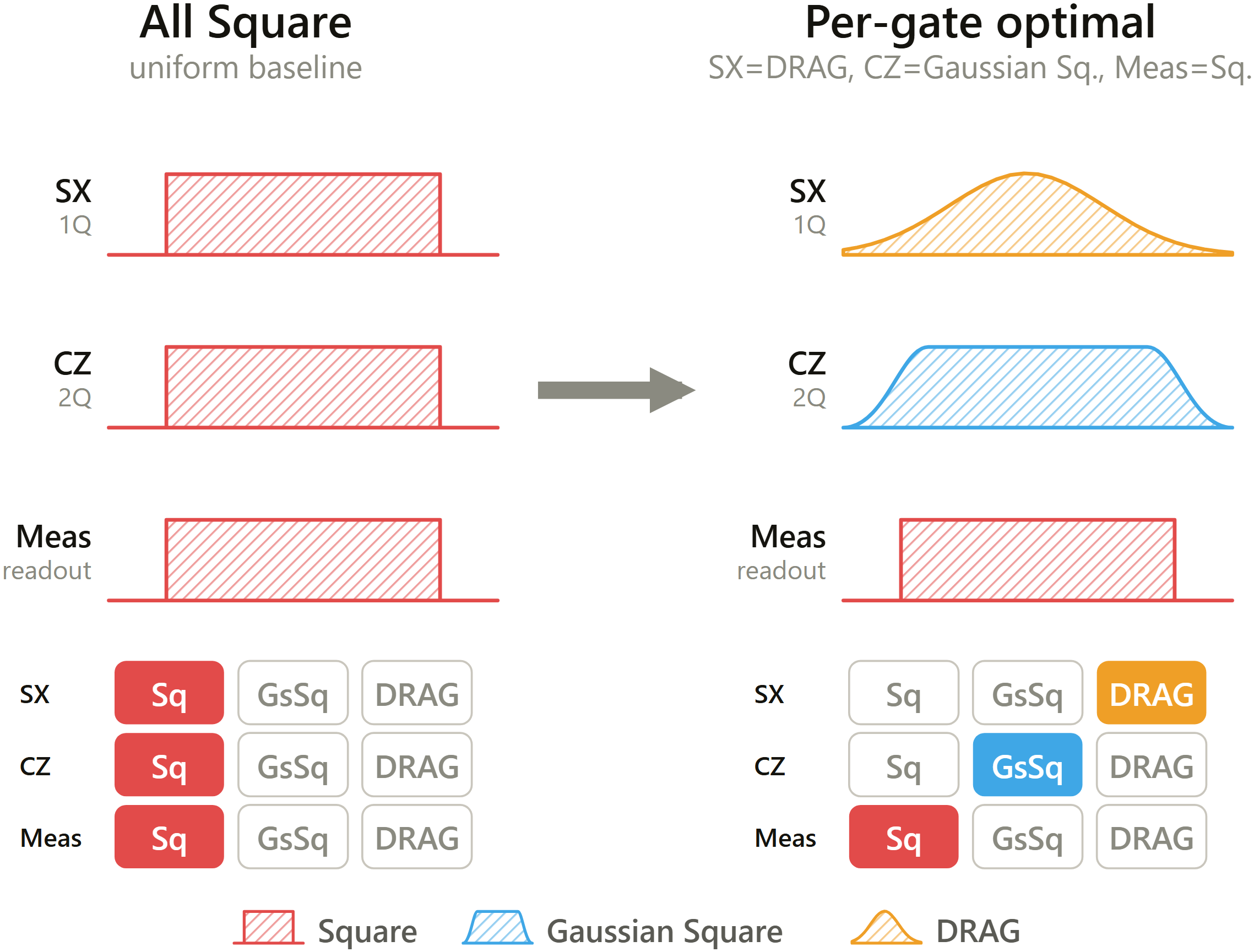}\label{fig:l3_ui}}\hspace{2mm}
\subfloat[Isolation experiment result.]{\includegraphics[height=6.4cm]{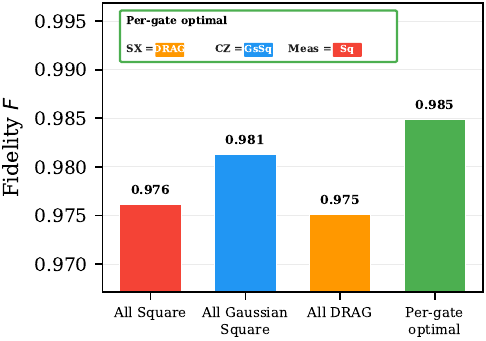}\label{fig:l3_result}}
\caption{L3 Pulse-shape Assignment. (a)~Illustrative rendering of the pulse-shape selector. The left panel shows the uniform All Square baseline (red hatched pulses for SX, CZ, Meas), and the right panel shows per-gate optimal (SX=DRAG orange, CZ=Gaussian Square blue, Meas=Square red). The 3$\times$3 selector matrix lets users pick a shape per gate type, with the coloured cell indicating the active choice. (b)~Per-gate optimal outperforms every uniform shape, achieving $F{=}98.49\%$, $+0.87\%$ over the All Square baseline (Fig.~\ref{fig:l2_result} endpoint).}
\label{fig:l3}
\end{figure*}

\subsection{L2 Sensitivity}
\label{sec:l2}

We fix L1 and L3 and perform an isolation sweep of L2's filtering parameters, identifying the physical mechanism through which qubit selection controls fidelity.

Two filter sliders (readout error $\tau_{\text{ro}}$ and 2Q gate error $\tau_{\text{2q}}$) progressively prune high-error edges from the coupling map (Fig.~\ref{fig:l2_ui}), yielding a filtered subgraph $G_\tau$. Filtering does not change the best achievable fidelity but raises the worst-case by excluding low-quality paths. We apply filters cumulatively and measure fidelity on the best and worst surviving paths ($n{=}60$, two input states, three repeats).

\begin{table}[t]
\centering
\caption{Cumulative L2 Filtering on IBM Torino}
\label{tab:e1g}
\begin{tabular}{lrcr}
\toprule
Filter & Paths & Worst $F$ & Band \\
\midrule
Baseline & 184 & 0.867 & 11.8\% \\
+$\tau_{\text{ro}}{<}5\%$ & 67 & 0.935 & 5.0\% \\
+$\tau_{\text{2q}}{<}1\%$ & 63 & 0.952 & 3.3\% \\
+$\tau_{\text{2q}}{<}0.5\%$ & 48 & 0.961 & 2.4\% \\
+$\tau_{\text{2q}}{<}0.3\%$ & 27 & 0.966 & \textbf{1.9\%} \\
\bottomrule
\end{tabular}
\\[2pt]
{\footnotesize\raggedright The ``Paths'' column counts the three-qubit candidate paths admitted by the cumulative threshold filter; the underlying coupling-graph edge count is shown in Fig.~\ref{fig:l2_ui}. The baseline filter requires $T_1{\geq}30\,\mu$s, $T_2{\geq}15\,\mu$s, $\tau_{1q}{<}1\%$, $\tau_{2q}{<}10\%$, admitting 184 paths. Best $F = 0.985$ throughout.\par}
\end{table}

As shown in Table~\ref{tab:e1g} and Fig.~\ref{fig:l2_result}, the readout error threshold $\tau_{\text{ro}}$ provides the largest single improvement ($F_{\text{worst}}$ goes from $0.867$ to $0.935$), and subsequent tightening of the 2Q gate threshold $\tau_{\text{2q}}$ closes the remaining gap.
At the strictest filtering, the fidelity band collapses to
$1.9\%$, and any surviving path achieves near-optimal
fidelity.


\subsection{L3 Sensitivity}
\label{sec:l3}

We fix L1 and L2 and vary L3's pulse-shape assignment in isolation.

QuBridge displays the circuit alongside its pulse-shape annotation (Fig.~\ref{fig:l3_ui}).

The effect of pulse-shape on fidelity is modeled as an analytical gate-level proxy rather than through pulse-level Hamiltonian simulation. Shape-dependent error rates, derived from established physical models~\cite{gambetta2011leakage}, are applied as Pauli-Lindblad rates in the gate-level noise model, enabling efficient evaluation without solving the Schr\"{o}dinger equation at the pulse level.

A teleportation circuit uses three gate types, single-qubit rotations (SX/X), two-qubit entangling gates (CZ), and measurement, each with different spectral characteristics. Rather than applying a uniform pulse-shape to all gates, we test whether \emph{per-gate shape assignment} improves fidelity. We compare three uniform-shape configurations (all gates set to Square, Gaussian Square, or DRAG) against a per-gate optimal assignment (SX$=$DRAG, CZ$=$Gaussian Square, Meas$=$Square), selected by matching each gate type to the shape that minimizes its dominant error channel (Fig.~\ref{fig:l3_result}).

The per-gate optimal assignment achieves $F{=}98.49\%$ (SX=DRAG, CZ=Gaussian Sq., Meas=Sq.), outperforming every uniform shape by at least $+0.36\%$ (Fig.~\ref{fig:l3_result}).
The improvement arises because different gate types have
different dominant error channels (single-qubit rotations favor
DRAG, two-qubit gates favor Gaussian Square, measurement is
shape-insensitive), and no uniform assignment can match all
three simultaneously.

L3's isolated effect ($+0.87\%$ over All Square) provides a context-fixed reference for the marginal effect observed in \S\ref{sec:trace}, where it appears alongside L2's larger contribution. Per-layer decomposition makes this contextualization observable.

\subsection{Error-Detected Teleportation}
\label{sec:logical}

We apply the same decomposition framework to error-detected teleportation, testing whether per-layer fidelity decomposition remains valid under encoding.
Encoding changes the decision landscape at each
layer. At L2, qubit selection shifts from
threshold-filtered linear paths (physical) to VF2 subgraph
matching (error-detected) with a fixed random seed
(seed=42) for deterministic tie-breaking among layouts of
equivalent noise score, since the six-qubit
$[[2, 1, 1]]$-encoded circuit requires a connected subgraph
that minimizes routing overhead.
\begin{figure}[!t]
\centering
\includegraphics[width=\columnwidth]{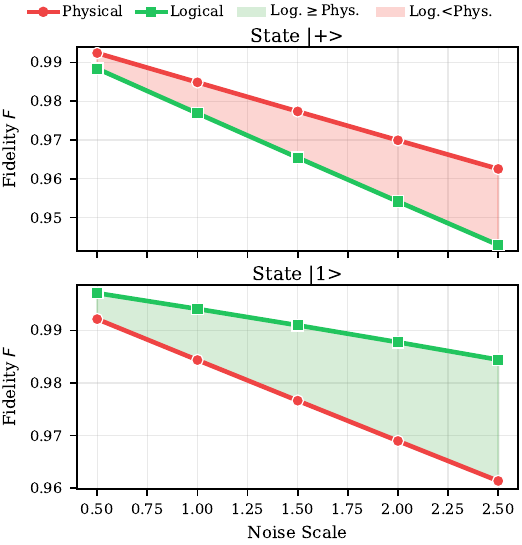}
\caption{Teleportation fidelity under increasing noise. For
phase-sensitive $\ket{+}$, logical fidelity stays below
physical because the $[[2,1,1]]$ code cannot detect $Z$
errors. For $T_1$-decay-sensitive $\ket{1}$, conditional
logical fidelity (after syndrome post-selection on
$\{00, 11\}$) exceeds physical across the tested noise
range as parity-changing events are removed, with the
conditional gap widening as noise grows. Table~\ref{tab:logical}
reports the corresponding acceptance rates.}
\label{fig:e3d}
\end{figure}

\begin{table}[t]
\centering
\caption{Teleportation Fidelity and Syndrome Acceptance Across Noise Scales}
\label{tab:logical}
\setlength{\tabcolsep}{3pt}
\begin{tabular}{lcccccc}
\toprule
 & \multicolumn{3}{c}{$\ket{+}$ (phase)} & \multicolumn{3}{c}{$\ket{1}$ ($T_1$ decay)} \\
\cmidrule(lr){2-4}\cmidrule(lr){5-7}
Noise & Phys.\ $F$ & Log.\ $F$ & Accept & Phys.\ $F$ & Log.\ $F$ & Accept \\
\midrule
0.5 & 0.9924 & 0.9884 & 0.9623 & 0.9922 & 0.9971 & 0.9613 \\
1.0 & 0.9849 & 0.9769 & 0.9263 & 0.9844 & 0.9941 & 0.9244 \\
1.5 & 0.9774 & 0.9655 & 0.8919 & 0.9767 & 0.9910 & 0.8892 \\
2.0 & 0.9699 & 0.9541 & 0.8590 & 0.9690 & 0.9878 & 0.8555 \\
2.5 & 0.9625 & 0.9429 & 0.8275 & 0.9614 & 0.9844 & 0.8234 \\
\bottomrule
\end{tabular}
\\[2pt]
{\footnotesize\raggedright Noise scale multiplies the calibrated error rates ($\text{ns}{=}1.0$ matches device calibration). Accept = fraction of shots with syndrome $\in \{00, 11\}$. Logical $F$ is conditional on acceptance; physical acceptance is unity by construction. Sign of $\Delta F$ and throughput-adjusted comparison are discussed in Section~\ref{sec:logical}.\par}
\end{table}

Error-detected teleportation is state-dependent
(Table~\ref{tab:logical}, Fig.~\ref{fig:e3d}). For state $\ket{+}$,
where $Z$-error dominates and is undetectable by the
$[[2,1,1]]$ code, logical fidelity remains below physical
across all noise levels and the gap widens from $-0.0040$ at
noise scale $0.5$ to $-0.0196$ at noise scale $2.5$. For
state $\ket{1}$, where $T_1$ decay changes the
computational-basis parity of the data qubits and is thus
flagged by the syndrome, post-selection keeps conditional
logical fidelity above physical across the tested range,
with the conditional advantage growing from $+0.0049$ at
noise scale $0.5$ to $+0.0230$ at noise scale $2.5$.
However, acceptance drops from $96.13\%$ to $82.34\%$ over
the same range, so the throughput-adjusted quantity
$\text{Log } F \times \text{Accept}$ favors physical
teleportation for state $\ket{1}$ as well, and a fortiori
for state $\ket{+}$ where conditional logical fidelity is
already below physical.
This state-dependent interaction between encoding (L1) and
layout optimization (L2) is what the per-layer framework
surfaces. Neither decision considered in isolation predicts
when encoding pays off.



\section{Discussion and Conclusion}
\label{sec:conclusion}

We presented QuBridge, a pipeline analysis tool that supports
three diagnostic tasks for quantum execution pipelines.
Practitioners can identify which layer is responsible for a
fidelity regression, estimate whether an upstream improvement
survives downstream transformations, and compare physical and
encoded executions under a common decision workflow.

Three findings change how practitioners should interpret
fidelity regressions. First, qubit selection acts as a
worst-case guarantee rather than a peak-performance lever.
Layout choice determines the floor of achievable fidelity,
not its ceiling, and this separation is invisible under
end-to-end measurement alone. Second, a layer's measured
contribution is context-dependent. The same pulse-shape
optimization yields different attributed gains depending on
whether upstream layout has been fixed, and per-layer
decomposition makes this dependence explicit. Third,
error-detection encoding is not uniformly beneficial, and
its conditional advantage emerges for input states whose
dominant error channel is detectable by the chosen code.

Within the small teleportation circuits evaluated here, L2 and
L3 are effectively independent because compact placements
within a local qubit cluster keep routing overhead small and
layout-independent. For larger circuits whose
connectivity exceeds the device's local topology, SWAP routing
becomes unavoidable, so L2's mapping decision determines
circuit depth and thereby the decoherence accumulation that L3's
pulse-shape assignment must mitigate. This tightens the coupling
between layers and makes per-layer decomposition more valuable
rather than less.

Our evaluation relies on density-matrix simulation with static
noise models from IBM Torino calibration data, which provides
reproducibility but does not capture crosstalk, spatially
correlated errors, or hardware drift. All experiments target
IBM Heron architecture, and generalization to other
architectures and to larger circuits where L3's decoherence
contribution dominates remains to be validated. The conditional
logical fidelity reported in Table~\ref{tab:logical} is measured
against a noise-aware physical baseline. Both physical and logical
layouts were selected by noise score, so any observed advantage
represents the residual gain of encoding on top of an already
well-chosen layout rather than over an arbitrary placement, and
for phase-sensitive states such as $\ket{+}$ no net advantage is
observed.

Future work includes quantifying inter-layer coupling at
larger circuit scales, extending the framework to variational
algorithms, and validating against live hardware.

\section*{Acknowledgments}

This work was supported by JST Moonshot R\&D Grant Number JPMJMS226C. SN acknowledges support from the JST Moonshot R\&D Program Grant No.\ JPMJMS256G and a JSPS Overseas Research Fellowship. We acknowledge IBM Quantum for the publicly available calibration data and the Qiskit open-source framework. The QuBridge source code, calibration snapshot, and CSV data for all figures and tables are available at \url{https://github.com/ukinsama/qubridge-qce26} under the MIT license; the artifact reproduces all results without live IBM Quantum hardware access.

\bibliographystyle{IEEEtran}

\end{document}